\documentclass[pra,amsmath,aps,twocolumn,showpacs]{revtex4}

\usepackage{graphicx}
\usepackage{color}
\usepackage{marvosym}
\usepackage{wasysym}
\usepackage{pifont}
\usepackage{amssymb}

\renewcommand{\r}{{\bf r}}
\newcommand{\gee}{\tilde{g}}
\newcommand{\partder}[2]{\frac{\partial{#1}}{\partial{#2}}}
\newcommand{\h}{\mathcal{H}}

\begin{document}

\title{Stability and dynamics of vortex clusters in nonrotated Bose-Einstein
  condensates}
\author{V.~Pietil\"a,$^1$ M.~M\"ott\"onen,$^{1,2}$
   T.~Isoshima,$^3$ J.~A.~M.~Huhtam\"aki,$^1$ and
  S.~M.~M.~Virtanen$^1$ }
\affiliation{$^1$Laboratory of Physics, Helsinki University of
  Technology, P.O.~Box 4100,
FI-02015 TKK, Finland \\
$^2$Low Temperature Laboratory, Helsinki University of Technology,
  P.O.~Box 3500,
FI-02015 TKK, Finland \\
$^3$Department of Physics, Graduate School of Science, Kyoto
  University, Kyoto 606-8502,
  Japan}
\date{\today}

\begin{abstract}
We study stationary clusters of vortices and antivortices in dilute
pancake-shaped Bose-Einstein condensates confined in nonrotating harmonic 
traps. Previous theoretical results on the stability properties of these
topologically nontrivial excited states are seemingly contradicting. We 
clarify this situation by a systematic stability analysis. 
The energetic and dynamic stability of the clusters is determined from the
corresponding elementary excitation spectra obtained by solving the 
Bogoliubov equations. Furthermore, we study the temporal evolution of the
dynamically unstable clusters. The stability of the clusters and the 
characteristics of their destabilizing modes only depend on the effective
strength of the interactions between particles and the trap 
anisotropy. For certain values of these parameters, there exist several
dynamical instabilities, but we show that there are also regions in 
which some of the clusters are dynamically stable. Moreover, we observe that
the dynamical instability of the clusters does not always imply 
their structural instability, and that for some dynamically unstable states
annihilation of the vortices is followed by their regeneration, and 
revival of the cluster.

\end{abstract}

\hspace{5mm}


\maketitle

\section{Introduction}

Quantized vortices are topological defects in systems with a long-range
quantum phase coherence. They have been extensively investigated in 
different systems and branches of physics such as helium
superfluids~\cite{donnelly:1991,vollhardt:1990},
superconductors~\cite{parks:1969},
cosmology~\cite{vilenkin:1994,anderson:1975}, and
optics~\cite{swartzlander:1992}. Atomic Bose-Einstein condensates (BECs) are
extremely convenient systems to investigate the characteristics of quantized
vortices due to their experimental versatility; many properties of these
systems can be manipulated with lasers and external magnetic
fields. Furthermore, real-space imaging of BECs can be carried out by optical
{\it in situ} or time of flight measurements.

In addition to extensive theoretical analysis, vortices and vortex clusters
composed of several vortices with the same topological charge  have 
been realized and investigated experimentally in gaseous
BECs~\cite{matthews:1999,madison:1999,abo:2001,raman:2001}. These vortex
clusters are local 
minima of energy for rotated condensates and hence stable states. However,
recently there has been theoretical suggestions that other kinds of vortex 
clusters can be stationary and stable states for nonrotated
condensates~\cite{crasovan:2002,crasovan:2003}. These clusters typically
consist of vortices and antivortices in specific configurations such that the
various forces acting on the vortices exactly balance each other. Based on 
the seeming robustness under external perturbations of some of these
topological excited collective states, they could also be considered as 
solitonic states---indeed, they seem to be stabilized only by strong enough
nonlinearity of the condensate. The development of techniques for 
vortex creation in dilute BECs, and especially phase-imprinting
methods~\cite{andrelczyk:2001}, may enable direct construction of such vortex 
clusters in the future. Both from the experimental and theoretical points of
view, their stability properties is a central issue. 

These vortex structures have been investigated in the noninteracting
limit~\cite{birula:2000,crasovan:2002} as well as in interacting systems
\cite{crasovan:2002,crasovan:2003,zhou:2004,mottonen:2005}. In the previous
studies, three different stationary vortex cluster configurations
have been found in interacting pancake-shaped condensates. The so-called
vortex dipole and quadrupole states, shown in Figs.~1(a) and 1(c), were
originally introduced by Crasovan {\em et
  al.}~\cite{crasovan:2002,crasovan:2003}. The stationary vortex quadrupole
state was found to exist in both interacting and noninteracting condensates,
whereas the stationary vortex dipole state exists only in 
condensates interacting strongly enough. Furthermore, the authors of
Refs.~\cite{crasovan:2002,crasovan:2003} studied the stability against 
small external perturbations of these configurations by integrating the
Gross-Pitaevskii equation in real time, and concluded them to be stable 
and robust in the nonlinear regime.

\begin{figure}[h]
\includegraphics[scale=0.43]{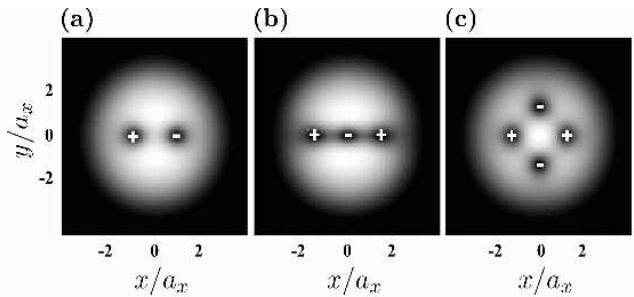}
\caption{\label{1comp_densities} Density profiles of the condensate for
  the stationary vortex dipole (a),
  tripole (b) and quadrupole (c) states. Vortices and
  antivortices are
  denoted by plus and minus signs at the vortex cores, respectively.
  The interaction strength parameter is $\tilde{g}=170$. The unit of length is
  the harmonic oscillator length in the $x$-direction.}
\end{figure}

Subsequently, based on the Bogoliubov quasiparticle spectra of the cluster 
states, the stability properties of the stationary clusters were 
studied by M\"ott\"onen {\em et al.}~\cite{mottonen:2005}. It was found that
at least for the parameter values studied, both the vortex dipole and vortex
quadrupole states are energetically and dynamically unstable. Effects of the
energetic instability can be neglected at low enough temperatures due to
vanishingly small dissipation, but the dynamical instabilities can prevent
these states from being long lived. In addition, another stationary cluster, a
vortex tripole, was introduced---see Fig.~1(b). 

In order to clarify the seeming contradiction between the results of
Refs.~\cite{crasovan:2002,crasovan:2003} and Ref.~\cite{mottonen:2005}, we 
present in this paper a systematic study of the stability properties of the
above-mentioned stationary vortex clusters as functions of total 
particle number and trap anisotropy. We find the stationary vortex cluster
states by directly minimizing an error functional for the 
Gross-Pitaevskii equation, and compute the corresponding instability modes,
{\em i.e.}, the quasiparticle excitations responsible for the 
dynamical instability, from the Bogoliubov equations. Finally, we investigate
the nature of the instability modes by computing the time 
development of slightly perturbed cluster states. It is found that the
characteristics of the instability modes only depend on the effective 
strength of the particle interactions and the trap anisotropy. Some dynamical
instability modes do not destroy the cluster structure and the 
condensate is therefore referred to as structurally stable, whereas some lead
to annihilation and subsequent revival of the vortices in the cluster.

\section{Mean field theory}\label{sec2}

The dynamics of weakly interacting gaseous Bose-Einstein condensates in the
zero-temperature limit is accurately
described by the Gross-Pitaevskii (GP) equation
\begin{equation}
\label{gp}
i\hbar\frac{\partial}{\partial t}\psi(\r,t) = \h\psi(\r,t),
\end{equation}
where the nonlinear operator $\h = \h[\psi]$ acting on the condensate order
parameter $\psi$ is given by
\begin{equation}
\label{1compHami}
\mathcal{H}[\psi] =
-\frac{\hbar^2}{2m}\nabla^2+V_{\mathrm{tr}}+g|\psi|^2.
\end{equation}
The interaction parameter~$g$ is determined by the $s$-wave scattering
length~$a$ as  $g = 4\pi\hbar^2 a/m$. Above,~$m$ is the atomic mass, and
the external trapping potential is denoted by~$V_{\textrm{tr}}(\r)$. The order
parameter is normalized according to $\int |\psi|^2\textrm{d}\r =
N$, where~$N$ is the total number of atoms in the condensate. The total energy
of the condensate can be calculated as
\begin{equation}
\label{e_func}
E[\psi] = \int
\textrm{d}\r\,\,\psi^{*}(\r)\bigg[-\frac{\hbar^2}{2m}\nabla^2+V_{\mathrm{tr}}(\r)  
+ \frac{g}{2}|\psi(\r)|^2\bigg]\psi(\r).
\end{equation}

The stationary states of the condensate with an eigenvalue~$\mu$ are solutions
to the GP equation of the form $\psi(\r,t)=e^{-i\mu
  t/\hbar}\psi(\r)$. Energetic and dynamic stability of a given stationary
state can be inferred from the corresponding excitation spectrum given 
by the Bogoliubov equations
\begin{equation}
\label{bogoliubov_eq}
\begin{pmatrix}
\mathcal{L}(\r) & g[\psi(\r)]^2 \\
-g[\psi^{*}(\r)]^2 & -\mathcal{L}(\r)
\end{pmatrix}
\begin{pmatrix}
u_q(\r) \\ v_q(\r)
\end{pmatrix}
= \hbar\omega_q
\begin{pmatrix} u_q(\r) \\ v_q(\r)
\end{pmatrix},
\end{equation}
where $\mathcal{L}=\mathcal{H}+g|\psi|^2-\mu$, the quasiparticle amplitudes
are denoted by $u_q(\r)$ and $v_q(\r)$, and $\omega_q$ is the 
eigenfrequency corresponding to the quasiparticle state with the index~$q$.

If the quasiparticle spectrum contains excitations with positive norm $\int
\textrm{d}\r\, (|u_q|^2-|v_q|^2) > 0$ but negative eigenfrequency
$\omega_q$, the corresponding stationary state is energetically unstable,
whereas eigenfrequencies with nonzero imaginary part indicate
dynamical instability. The occupation of dynamical instability modes increases
exponentially in time, and hence small perturbations of a
dynamically unstable stationary state typically result in large modifications
in its structure. 

In pancake-shaped traps, the system in question has a~$SO(2)$ rotation
symmetry. However, the vortex cluster states shown in 
Fig.~\ref{1comp_densities} are not rotationally invariant. 
Thus, according to the Goldstone theorem~\cite{goldstone:1961,forster:1975},
there should 
exist collective low-frequency modes which tend to restore the broken 
symmetry by rigidly rotating the cluster states. These modes are intrinsic to
the vortex cluster states, since they are absent, {\em e.g.}, for 
an axially symmetric single vortex state.

Existence of a symmetry breaking induced long-lived collective mode is
manifested as a~$1/k^2$ divergence in the response function of a 
symmetry restoring variable~$F$~\cite{forster:1975}. Let us
consider operators 
of the form \cite{pitaevskii:2003} 
\begin{align}
F = &\sum_q\int\textrm{d}\r\,f(\r)\bigg[\big(\psi^*(\r)u_q(\r) +
\psi(\r)v_q(\r)\big)b_q^{} e^{-i\omega_q^{}t}
\nonumber \\
&+\big(\psi(\r)u_q^*(\r) +
\psi^*(\r)v_q^*(\r)\big)b_q^{\dagger}e^{i\omega_q^{}t}\bigg], 
\end{align}
where~$b_q^{\dagger}$ and~$b_q$ are the creation and annihilation operators
for a quasiparticle with index~$q$ and~$f(\r)$ is a complex 
function. In the presence of dynamical instability modes, we may generalize
the result presented Ref.~\cite{pitaevskii:2003} to obtain the 
response function as
\begin{align}
\chi_{\scriptscriptstyle F}^{}(\omega,t) =
-\frac{1}{\hbar}\sum_q
&\bigg[\frac{A_q}{\omega-\omega_{q,0}^{}+i\eta}e^{-2\textrm{Im}\{\omega_q\}t}
\nonumber \\
\label{response}
&- \frac{B_q}{\omega+\omega_{q,0}^{}+i\eta}e^{2\textrm{Im}\{\omega_q\}t}\bigg],
\end{align}
with
$
A_q = \big|\int\textrm{d}\r\,f^*(\r)\big(\psi^*(\r)u_q(\r) +
\psi(\r)v_q(\r)\big)\big|^2$ and 
$B_q = \big|\int\textrm{d}\r\,f(\r)\big(\psi^*(\r)u_q(\r) +
\psi(\r)v_q(\r)\big)\big|^2$.
The Bohr frequency~$\omega_{q,0}^{}$ is given by $\omega_{q,0}^{} =
(E_q-E_0)/\hbar$, where~$E_q=\textrm{Re}\{\hbar\omega_q^{}\}$ is the energy of
the  excitation~$q$ and~$E_0$ is the ground state energy.  
From Eq.~\eqref{response} we observe that a necessary condition for the
response 
function $\chi_{\scriptscriptstyle F}^{}$ to diverge as~$1/\omega$, {\em
  i.e.},  as $1/k^2$, is the existence of excitations with
$\textrm{Re}\{\omega_q\} =0$. In Sec.~\ref{sec4}, we show that 
there indeed exist a zero-energy instability mode for all stationary vortex
clusters and we therefore identify them as the Goldstone modes corresponding
to broken rotational symmetry. 

\section{Computational methods}\label{sec3}

In Ref.~\cite{mottonen:2005}, the stationary vortex clusters considered were
found to be energetically unstable. Thus, they cannot be found with 
the usual methods based on energy minimization. We employ the method
introduced in Ref.~\cite{mottonen:2005}, in which stationary solutions of 
the GP~equation~\eqref{gp} are found by minimizing the error functional
$F[\psi,\mu] = \int \textrm{d}\r\,|(\mathcal{H}[\psi] - \mu)\psi(\r)|^2$. The
chemical potential which minimizes the error functional is given by the
familiar expression $\mu =
\int\textrm{d}\r\,\psi^*(\r)\mathcal{H}[\psi]\psi(\r)/N$. Using the functional
gradient 
\begin{align}
\label{grad} \frac{\delta F[\psi,\mu]}{\delta\psi^*} =&
\big[(\mathcal{H}[\psi]-\mu)^2+2g\mathrm{Re}\{\psi^*(\r)(\mathcal{H}[\psi]
\nonumber  \\ 
& -\mu)\psi(\r)\}\big]\psi(\r),
\end{align}
one obtains the stationary solutions, {\em e.g.}, by the method of steepest
descent or by directly solving the equation $\delta F/\delta\psi^* = 0$.

For computational simplicity, we consider a pancake-shaped system in the
harmonic potential
\begin{equation}
V_{\mathrm{tr}}(\r) = \frac{1}{2}m\left(\omega_x^2x^2 + \omega_y^2y^2 +
  \omega_z^2z^2\right),
\end{equation}
with $\omega_x,\omega_y \ll \omega_z$. Especially, we assume the
$z$-confinement to be tight enough, such that the low energy solutions of the 
GP equation and the Bogoliubov equations can be taken to be of the form
$\sigma(z)=1/\sqrt{2\pi a_z}e^{-z^2/(2a_z^2)}$ in the $z$-direction. 
If the harmonic oscillator length $a_z=\sqrt{\hbar/(m\omega_z)}$ in the axial
direction is much larger than the scattering length $a$, the 
condensate can be described by the usual GP equation with
$g=4\pi\hbar^2a/m$~\cite{petrov:2000}. Substituting the ansatz $\psi(\r,t) = 
\psi_{\textrm{\tiny 2D}}^{}(x,y,t)\sigma(z)$ into Eq.~\eqref{gp}, one obtains
the effectively two-dimensional (2D) Gross-Pitaevskii equation in 
the dimensionless form
\begin{equation}
\label{dimensionless}
i\partder{}{\tilde{t}}\tilde{\psi} = \frac{1}{2}\left[-\tilde{\nabla}^2 +
(\tilde{x}^2+\lambda^2 \tilde{y}^2) + \tilde{g}|\tilde{\psi}|^2 +
\frac{\omega_z}{\omega_x}\right]\tilde{\psi},
\end{equation}
where $\tilde{\psi}(\tilde{x},\tilde{y},\tilde{t}) = a_x\psi_{\textrm{\tiny
    2D}}^{}(x,y,t)$ and $a_x=\sqrt{\hbar/(m\omega_x)}$.
The dimensionless quantities denoted by the tilde are obtained from the
    original dimensional ones by scaling the length by~$a_x$, time by
$1/\omega_x$, and energy by $\hbar\omega_x$. The dimensionless constant
    $\lambda = \omega_y/\omega_x$ characterizes trap anisotropy in the
$xy$-plane. The normalization condition
    $\int\textrm{d}\tilde{x}\textrm{d}\tilde{y}\,|\tilde{\psi}|^2 =1$ implies
    the effective dimensionless 
2D coupling constant to be $\tilde{g} =
    4\sqrt{\pi}Na/a_z$. Equation~\eqref{dimensionless} implies that the only
    relevant parameters in the 
problem are the coupling constant~$\gee$ and the trap
    anisotropy~$\lambda$. Thus it suffices to investigate the cluster states
    as functions of 
these parameters.

Based on the minimization of the error functional, we have computed stationary
cluster states for different values of the interaction strength 
$\gee$ and trap asymmetry $\lambda$. To obtain a proper starting point for the
minimization of the error functional $F[\psi,\mu]$, we first 
minimize the energy of ansatz wave functions for which the condensate phase is
fixed. The found stationary states were computed to the relative 
final accuracy $F[\psi_s,\mu]/(N\mu^2) < 10^{-15}$, which confirms that these
states were, indeed, very accurate stationary states. The 
Bogoliubov equations were then solved for these states, identifying possible
excitation modes with non-real eigenfrequencies. Furthermore, we 
studied the nature of the found instability modes by solving the condensate
temporal evolution for initial, slightly perturbed states of the form 
\begin{equation}
\label{psi_excit}
\psi(\r,0) = \psi_s(\r) + \kappa_q[u_q(\r) + v_q^*(\r)],
\end{equation}
where $\psi_s(\r)$ is the condensate order parameter for a stationary state,
and $u_q(\r)$, $v_q(\r)$ are the quasiparticle amplitudes of an 
instability mode. The constant $\kappa_q$ describes the initial population of
the instability mode---by using  the normalization $\int d\r\, 
(|u_q|^2+|v_q|^2)=1$ for the instability modes, we typically chose $\kappa_q
\sim 10^{-2}\,\textrm{--}\,10^{-1}$ corresponding to less than one 
percent of the atoms in the unstable mode. We also point out that contrary to
the real-frequency modes, the energy of the perturbed stationary 
state $\psi(\r,0)$ in Eq.~\eqref{psi_excit} is up to second order in
$\kappa_q$ equal to the energy of the stationary state $\psi_s(\r)$ if the 
perturbation corresponds to a dynamical instability mode. Furthermore, our
numerical computations have showed that in the presence of several 
instability modes in the spectrum of the state, it is typically sufficient to
consider only the excitation with the largest 
$|\textrm{Im}\{\omega_q\}|$. This is due to the fact that quasiparticle
interactions necessarily excite also the fastest growing imaginary mode 
which seems to always dominate the condensate dynamics. The temporal evolution
of the perturbed state was solved numerically from the 
time-dependent GP~equation using a combination of a split-operator method and
an implicit Crank-Nicolson scheme.

\section{Results}\label{sec4}

\begin{figure}[!h]
\includegraphics[scale=0.85]{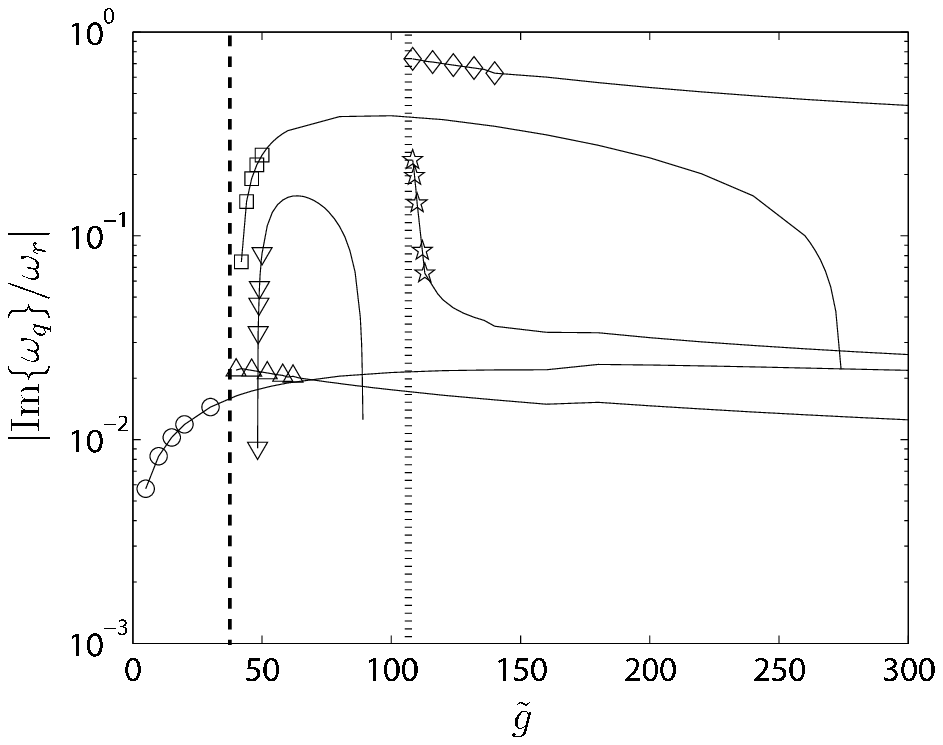}
\caption{\label{im_parts}  Absolute values of the imaginary parts of the
  eigenfrequencies corresponding to instability modes as functions of the
interaction strength $\gee$ for stationary dipole, tripole and quadrupole
clusters confined in rotationally symmetric trap, with $\lambda=1$. 
Curves corresponding to the vortex dipole are marked with~$\scriptstyle
\bigtriangledown$ (decay mode) and~$\scriptstyle \bigtriangleup$ 
(rotational mode), the vortex tripole with~$\scriptstyle \diamondsuit$ (decay
mode) and~\ding{73} (rotational mode), and the vortex quadrupole 
with~$\scriptstyle \square$ (decay mode) and~$\scriptstyle \ocircle$
(rotational mode). The dashed and the dotted line indicate the lower limits 
of the interaction strength for which stationary vortex dipole and tripole
were found to exist, respectively. } 
\end{figure}

The vortex cluster states were analyzed by varying the interaction parameter
in the range $0\leq \gee\leq 300$ for the rotationally symmetric 
trap with $\lambda=1$, and by separately varying the anisotropy parameter
$\lambda$ for fixed interaction parameter values. Specifically, we 
studied the existence of the stationary vortex clusters for $\tilde{g}=170$
(which was used in the previous studies in Ref.~\cite{mottonen:2005}) 
and the existence of certain instability modes for a wide range of 
the interaction parameter. Excitation spectra for all the stationary clusters
considered contained negative energy excitations with positive 
norm, implying these configurations to be energetically unstable. On the other
hand, dynamical instability of the clusters turned out to be a 
more subtle issue. The imaginary parts of the dynamical instability modes of
the clusters in rotationally symmetric traps are presented in 
Fig.~\ref{im_parts} as functions of the interaction strength
$\gee$. Inspecting the temporal evolution of the cluster states in the
presence of excited instability modes, we observe that there are
two qualitatively different dynamical instability modes for each cluster; one
which turns out to lead to decay of the cluster and one which tends to rotate
the cluster rigidly. We identify the latter modes as the Goldstone modes
discussed in Sec.~\ref{sec2}, since they have zero energy and they vanish if
the trap becomes sufficiently anisotropic. 

\subsection{Vortex dipole}

\begin{figure}[!h]
\includegraphics[scale=0.40]{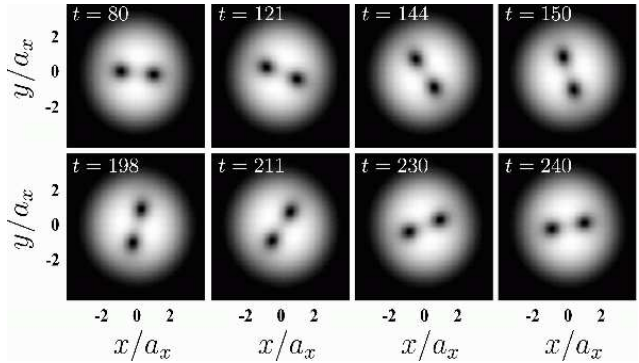}
\caption{\label{dipole_rot} Temporal evolution of the slightly perturbed
  stationary vortex
  dipole state for $\tilde{g} = 160$ and $\lambda = 1$. The interaction
  strength $\gee = 160$ is in the region where only the rotational mode
  exists. Time is denoted by $t$
  and it is given in the units of
  $1/\omega_x$. The structure of the cluster remains very close to the
  original stationary configuration.}
\end{figure}

For a rotationally symmetric trap with $\lambda=1$, the stationary vortex
dipole state was found to exist only for interaction strengths 
$\gee\gtrsim 42$, {\em i.e.}, certain amount of nonlinearity is required for
the existence of the stationary dipole configuration. This agrees 
with the observations presented in Ref.~\cite{crasovan:2003}. For the vortex
dipole, there exists two dynamical instability modes: one for all 
interaction strength values $\gee\simeq 42$--$300$ considered (marked with
$\scriptstyle \bigtriangleup$ in Fig.~\ref{im_parts}), and one that 
occurs only in the range $\gee\simeq 50$--$80$ (marked with $\scriptstyle
\bigtriangledown$ in Fig.~\ref{im_parts}). The latter mode dominates 
in the region $\gee\simeq 50$--$80$ due to its larger
$|\textrm{Im}\{\omega_q\}|$. Time-development of slightly perturbed vortex
dipole states 
shows that the mode existing outside the region $\gee\simeq 50$--$80$
corresponds very accurately to rigid rotation of the dipole configuration,
as shown in Fig.~\ref{dipole_rot}. The structure of the dipole remains intact
in this rotation, and hence the stationary dipole state can be
considered structurally stable, although it is dynamically unstable in this
region. Thus dynamical instability does not necessarily imply the
cluster structure to be unstable.

\begin{figure}[!h]
\includegraphics[scale=0.40]{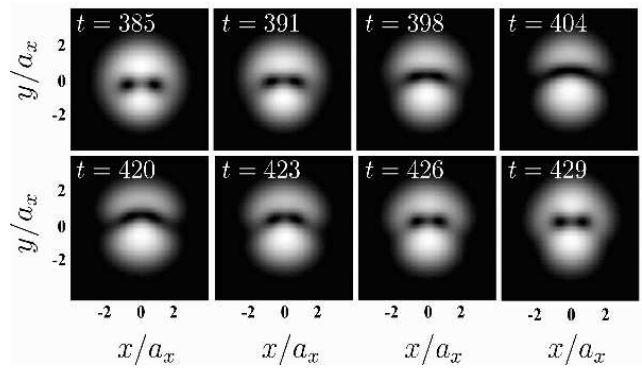}
\caption{\label{dipole_ann} Collapse and revival of the slightly perturbed
  stationary vortex dipole state for $\tilde{g} = 60$ and $\lambda = 1$. It
  should be noted that apart from oscillations of the vortex locations,
  the vortex dipole state remains essentially intact 
  for a long time before it starts to decay. The vortices have opposite
  topological charges after the revival of the dipole state. }
\end{figure}

The nature of the instability mode dominating vortex dipole dynamics in the
region $\gee\simeq 50$--$80$ is very different from the rotational
mode. Temporal evolution of a slightly perturbed dipole state in this regime
is shown in Fig.~\ref{dipole_ann}. The dominant mode renders the
vortices of the dipole first to annihilate each other, but eventually the
dipole configuration reappears from the vortex-free state. For $\gee
\simeq 50$, the density distribution of this revived dipole configuration is
almost the same as in the initial state, but the vorticity is
changed such that the vortex becomes an antivortex and {\em vice
  versa}. Similar 
behavior has been observed in the numerical simulations of light
propagation in a graded-index medium, where the vortex dipole nested in a
light beam undergoes periodical collapse and revival along the
direction of the light propagation~\cite{molina-terriza:2001}. The
collapse and revival of the dipole configuration continues periodically
for $\gee \simeq 50$, whereas for increasing interaction strength,
nonlinearity gradually prevents the topological excitation from recombining
and eventually the whole excitation disappears for $\gee \gtrsim 80$. 

In asymmetric traps, the stationary vortex dipole state corresponding to
interaction parameter $\gee=170$ was found to exist in the range 
$\lambda\simeq 0.9$--$1.5$. Interestingly, we found that even the tiny amount
of asymmetry corresponding to $\lambda=1.005$ is sufficient to 
eliminate the rotational instability mode for all values of $\gee$. This is
consistent with the fact that in an asymmetric trap, the system no longer has 
the $SO(2)$ symmetry and the Goldstone mode should vanish. Thus by
using a slightly asymmetric trap the stationary vortex dipole state 
can be made fully dynamically stable in ranges $\gee\simeq 42$--$50$ and
$\gee\gtrsim 80$. Experimentally, the transfer between the different 
stability and instability regimes can be accomplished by adjusting the total
particle number and the trapping frequencies.

\subsection{Vortex tripole}

\begin{figure}[!h]
\includegraphics[scale=0.38]{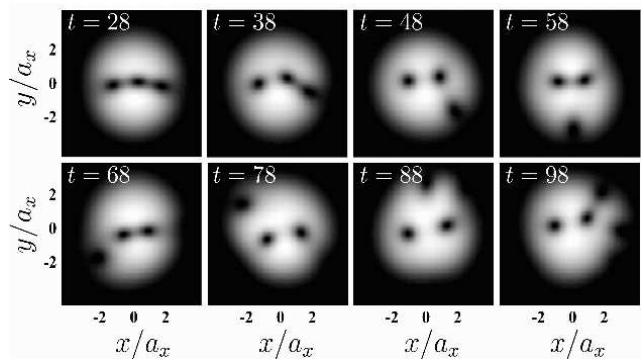}
\caption{\label{vt_td} Temporal evolution of the perturbed stationary vortex
  tripole state for $\tilde{g} = 160$ and $\lambda = 1$. The vortex tripole
  starts to decay almost 
  immediately to the vortex dipole state due to the strong dynamical
  instability. } 
\end{figure}

The vortex tripole state exists only in rather strongly interacting
condensates: in symmetric traps with $\lambda=1$, the stationary vortex
tripole state was found to exist only for $\gee \gtrsim 108$. The stationary
vortex tripole has always two dynamical instability modes, as shown
in Fig.~\ref{im_parts}. The real part of the eigenfrequency vanishes for both
of these modes, but the imaginary part of the dominating mode is
roughly an order of magnitude larger than that of the other one. The decay of
the tripole configuration corresponding to the dominating
instability mode is shown in Fig.~\ref{vt_td}. Under small perturbations, one
of the outermost vortices starts to drift out of the condensate
and eventually it reaches the surface of the cloud and excites surface modes
which can be observed in the last panel in Fig.~\ref{vt_td}. The
remaining two vortices form a topologically neutral vortex dipole.
Furthermore, the stationary vortex tripole exists as a stationary state only
for $\gee \gtrsim 108$ which lies in the structurally stable region of the
stationary vortex dipole. In the numerical simulations, the vortex
tripole takes a time of order $1/\omega_x$ for the population $\kappa_q\sim
10^{-2}\textrm{--}10^{-1}$ of the
excitation, before it starts to notably decay. Thus the decay of the vortex
tripole to the vortex dipole is slow
enough to be experimentally observable. The nature of the other
instability mode, to which we refer as the slow mode, is more difficult to
find out since this mode never seemed to dominate the temporal
evolution. Numerical simulations showed that the slow mode tends to rotate the
condensate, but the evolution of this mode excites also the
dominant instability mode leading to the decay of tripole cluster to dipole
cluster.

In anisotropic traps, the vortex tripole state turned out to be stationary in
the same range $\lambda\simeq 0.9$--$1.5$ for $g\approx 170$ as
the vortex dipole. Anisotropy of the trap corresponding to $\lambda\geq 1.01$
removes the slow instability mode from the spectrum, but the
dominating mode always persists. This observation together with the
characteristics of the initial decay dynamics suggests that the slow mode is
analogous to the rotational mode of the vortex dipole and quadrupole. Since
the dominant mode persists in an asymmetric trap, the vortex tripole
state cannot be made dynamically stable by tuning the interaction strength or
the trap anisotropy.

\subsection{Vortex quadrupole}

The vortex quadrupole was found to be stationary for all interaction strengths
for $\lambda = 1$. These results are in agreement with the
previously reported calculations of Crasovan {\em et
  al.}~\cite{crasovan:2002}, in which stationary vortex quadrupoles were found
to exist both in interacting and noninteracting condensates. In this respect,
the quadrupole configuration differs essentially from the dipole
and tripole clusters. The dynamical instability modes of the quadrupole
cluster resemble those of the vortex dipole: The stationary vortex
quadrupole state in rotationally symmetric trap has two instability modes, one
which exist for all values of the interaction parameter, and one
existing only in the range $\gee\simeq 50$--$280$, where it dominates the
condensate dynamics, see Fig.~\ref{im_parts}. The mode which exists
for all $\gee$ has $\textrm{Re}\{\omega_q\}=0$, whereas the other mode has
energy $\textrm{Re}\{\hbar\omega_q\} \sim 10^{-1}\hbar\omega_r$.

\begin{figure}[!h]
\includegraphics[scale=0.35]{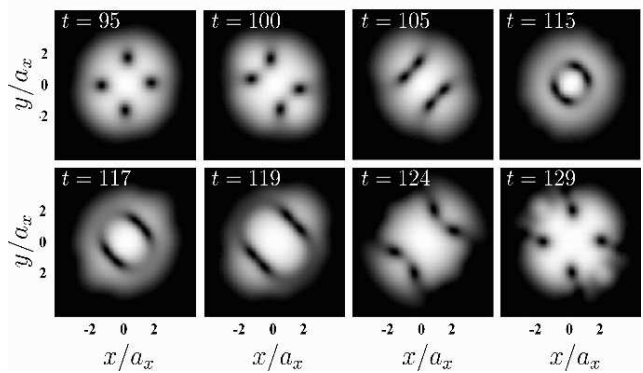}
\caption{\label{vq_td} Collapse and revival of the vortex quadrupole for
  $\tilde{g} = 260$ and $\lambda = 1$. In contrast to the vortex dipole
  state, the vortices have the same topological charge before and after the
  collapse and revival of the cluster. }
\end{figure}

In the region $\gee\simeq 50$--$280$, the dominant mode of the vortex
quadrupole drives the vortices to merge together and annihilate each
other, but eventually the quadrupole configuration reappear as shown in
Fig.~\ref{vq_td}. This mode also generates oscillations with increasing
amplitude such that the condensate stretches and shrinks as the
annihilation-revival cycles proceed. The  amplitude of these oscillations
increases gradually, and finally the vortices are driven out of the
condensate. In the regions $\gee \lesssim 60$ and $\gee\simeq 280$--$300$,
the existing instability mode is analogous to the dipole cluster mode which
corresponds to rigid rotation of the cluster, {\em i.e.}, 
the Goldstone mode. Thus the quadrupole
cluster is structurally stable in this regime. Furthermore, one observes from
Fig.~\ref{im_parts} that the vortex quadrupole tends to become
dynamically stable in the limit of noninteracting condensate.
It has, however, been shown that in the noninteracting case the persistent
current such as the vortex quadrupole is always structurally
unstable against perturbations in the external trap
parameters~\cite{garcia-ripoll:2001,comment}. Thus dynamical stability does not
necessarily imply stability against small perturbations in the trap parameters
and {\em vice versa}, dynamically unstable states can be structurally robust
with respect to small perturbations.

As a function of the trap anisotropy parameter, the quadrupole cluster
with $\gee=170$ was found to be stationary in the range $\lambda\simeq
0.9$--$1.2$. 
For increasing trap anisotropy parameter, the two vortices of the vortex
quadrupole in the direction of the tight confinement move away from
the center of the cloud, and eventually the state cannot be considered as a
vortex quadrupole. This behavior is due to the increase of the
buoyancy force in the direction of tight confinement and repulsive interaction
between vortices with same topological charge.

Analogously to the vortex dipole case, computations with different interaction
strength values $\gee$ showed that for the stationary vortex
quadrupole state, even the rather small amount of asymmetry $\lambda = 1.1$
is sufficient to remove the rotational instability mode from the
quasiparticle spectrum. Hence, also the vortex quadrupole state can be made
dynamically stable in the regimes $\gee \lesssim 60$ and
$\gee\gtrsim 280$.

\section{Conclusions}\label{sec6}

We have systematically studied the existence, stability, and dynamics of
stationary vortex clusters in dilute pancake-shaped BECs confined in
nonrotating harmonic traps. In contrast to the previous
investigations~\cite{crasovan:2002, crasovan:2003}, our approach utilizes the
Bogoliubov
equations to determine the quasiparticle spectrum which reveals unambiguously
not only the energetic and dynamic stability of a given state, but
also yields explicitly the quasiparticle amplitudes of the instability
modes. The nature of these modes was found out by computing the temporal
evolution of slightly perturbed stationary cluster states.

It was observed that the dipole, tripole and quadrupole clusters have various
regimes of dynamical instability, but in some of these
regimes the cluster configurations are structurally stable as they only tend
to rotate rigidly. On the other hand, the vortex annihilation modes
of the vortex dipole and quadrupole break the cluster structures altogether,
but if energy dissipation is negligible, the clusters can reappear
after a certain time. Furthermore, it was observed that even very small trap
anisotropies suffice to remove the rotational instability modes,
thus dynamically stabilizing the dipole and quadrupole clusters for suitable
values of the interaction parameter.

\begin{acknowledgments}

CSC-Scientific Computing Ltd (Espoo, Finland) is acknowledged for
computational resources and Academy of Finland for financial support through 
the Center of Excellence in Computational Nanoscience. MM and JH thank Finnish
Cultural Foundation, and MM and TI Vilho, Yrj\"o, and Kalle 
V\"ais\"al\"a Foundation for financial support.

\end{acknowledgments}

\bibliography{manuscript}

\end{document}